\documentclass[aps,prb,twocolumn,groupedaddress,showpacs]{revtex4-1}
\usepackage{times,amssymb,amsmath}
\usepackage[pdftex]{graphicx}
\renewcommand{\vec}[1]{\mbox{\boldmath$\mathrm{#1}$}}
\mathchardef\mhyphen="2D
\begin{document}
\title{Superatom molecular orbitals: a new type of long-lived electronic states}
\author{Y. Pavlyukh}
\email[]{yaroslav.pavlyukh@physik.uni-halle.de}
\author{J. Berakdar}
\affiliation{Institut f\"{u}r Physik, Martin-Luther-Universit\"{a}t
  Halle-Wittenberg, Heinrich-Damerow-Str. 4, 06120 Halle, Germany}\date{\today}
\begin{abstract}
We present \emph{ab initio} calculations of the quasiparticle decay times in a
Buckminsterfullerene based on the many-body perturbation theory.  A particularly lucid
representation arises when the broadening of the quasiparticle states is plotted in the
angular momentum ($\ell$) and energy ($\varepsilon$) coordinates. In this representation
the main spectroscopic features of the fullerene consist of two occupied nearly parabolic
bands, and delocalized plane-wave-like unoccupied states with a few long-lived electronic
states (the superatom molecular orbitals, SAMOs) embedded in the continuum of Fermi-liquid
states.  SAMOs have been recently uncovered experimentally by M.~Feng, J.~Zhao, and
H.~Petek [Science {\bf 320}, 359 (2008)] using scanning tunneling spectroscopy.  The
present calculations offer an explanation of their unusual stability and unveil their
long-lived nature making them good candidates for applications in the molecular
electronics. From the fundamental point of view these states illustrate a concept of the
Fock-space localization [B.~L.~Altshuler, Y.~Gefen, A.~Kamenev, and L.~S.~Levitov, Phys.
Rev. Lett. {\bf 78}, 2803 (1997)] with properties drastically different from the
Fermi-liquid excitations.
\end{abstract}
\maketitle
\begin{figure}[ht!]
\includegraphics[width=\columnwidth]{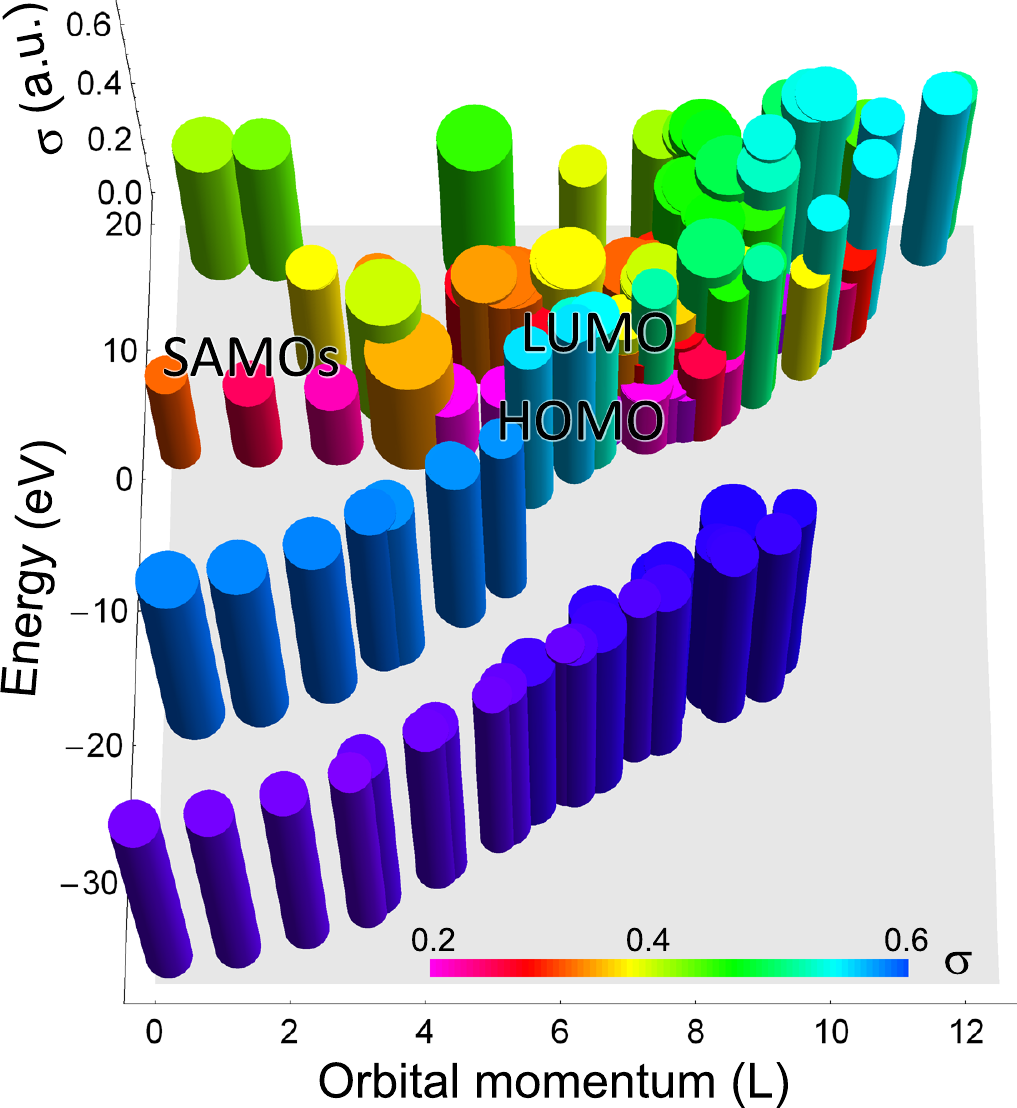}
\caption{Single-particle states of C$_{60}$ molecule. Each group of the degenerate states
  is represented by a cylinder and is characterized by the averaged orbital momentum
  ($x$-axis), the energy ($y$-axis), the energy uncertainty $\sigma$ (atomic units (a.u.),
  $z$-axis or colour coding) and the orbital momentum uncertainty ($\delta \ell$, the area
  of the cylinder base). The occupied valence states (as calculated fully
  \emph{ab-initio}) form two nearly parabolic bands that can be described analytically by
  $E_n(\ell)=E_n(0)+\frac{\ell(\ell+1)}{2R^2},\quad n=1,\,2$. This is a signature of the
  delocalized motion of the electrons on the spherical cage of the radius $R=3.57$~{\AA}
  (Ref.~\onlinecite{Lband}). SAMOs show comparable values of $\delta \ell$ and much
  smaller $\sigma$ and form a third band dissolving in the sea of other unoccupied
  states. Note that the weak dispersion of the SAMOs with $\ell$, as observed above, is
  due to their larger averaged radius $R$, as compared to the first two angular
  subbands. \label{fig:c60}}
\end{figure}

Superatom molecular orbitals (SAMOs) were recently discovered~\cite{Petek2008} as
universal characteristics of C$_{60}$ molecules and their aggregates. Further experiments
on endohedral systems~\cite{Petek2009} and calculations for a series of quasi-spherical
molecules~\cite{Lband} showed that SAMOs are associated with the whole system (rather then
with a particular atom), have a well defined spherical symmetry (Fig.~\ref{fig:c60},
angular momentum $\ell=0,1,2$, principal quantum number $n=3$) and are capable of forming
chemical bonds. Being markedly different from other unoccupied delocalized states they
hold a promise for unique applications in molecular electronics.  Hitherto theory was not
able to answer why these states are so resilient to the chemical environment and why their
character is not washed out by the hybridization: questions of critical importance for the
SAMO-mediated charge transport~\cite{Petek2010}.

To resolve these issues we report here results of quantum chemical calculations for
prototypical C$_{60}$ using recently developed self-energy
formalism~\cite{exp-set-up}. When a transport electron is injected into the unoccupied
SAMO (the single-particle ($1p$) state $\alpha$) the molecule undergoes a transition into
the mixed quantum state that decays in time. At later instances the decay is exponential
$\exp(-\vec\gamma_\alpha t)$ with $\vec \gamma_\alpha$ given by the imaginary part of the
on-shell self-energy $\vec \gamma_\alpha=\Im\vec{\Sigma}_{\alpha\alpha}(\epsilon)$
(Ref.~\onlinecite{CI-GW}). However, it is the initial stage of the decay that is
manifested in ultrafast nonequilibrium processes of technological relevance. We have
shown~\cite{exp-set-up} that the spectral function for a large class of relevant
electronic systems can be represented in the form
\begin{equation}
\vec A_\alpha(t)=\exp\left(-\vec \gamma_\alpha\frac{t^2}{t+\vec \tau_\alpha}\right),\label{eq:At}
\end{equation}
where the set-in time of the exponential decay is given by
$\vec\tau_\alpha=2\vec\gamma_\alpha/\vec{\sigma}^2_{\alpha\alpha}$. The spectral function
$\vec A_\alpha(t)$ has the following short and long time-limits:
\begin{eqnarray}
\frac{d}{dt}\vec A_\alpha(t)&\stackrel{t\rightarrow0}{\rightarrow}
&-\sigma^2_{\alpha\alpha}t\label{eq:shr-tm},\\
\vec A_\alpha(t)&\stackrel{t\rightarrow\infty}{\rightarrow}&e^{-\vec \gamma_\alpha t},\label{eq:lng-tm}
\end{eqnarray}
meaning that for short times ($t\ll\tau_\alpha$) we have the quadratic decay
\begin{equation}
 \vec A_\alpha(t)\sim 1-\vec{\sigma}^2_{\alpha\alpha}t^2/2.
\end{equation}

$\vec{\sigma}^2_{\alpha\alpha}$ is the central quantity for our theory. It can be
 determined as follows.  A simple substitution of asymptotic expansions into the Dyson
 equation leads to the exact relations between the spectral moments of the self-energy $\vec
 \Sigma(\omega)$ and of the spectral function $\vec A(\omega)$
 (Ref.~\onlinecite{Vogt2004}):
\begin{eqnarray}
\vec{M}^{(0)}=\vec{I}, &\quad&
\vec{\Sigma}_{\infty}=\vec{M}^{(1)}-\vec{\varepsilon},\label{eq:srule1}\\
\vec{\Sigma}^{(0)}&=&\vec{M}^{(2)}-[\vec{M}^{(1)}]^2,\label{eq:srule2}
\end{eqnarray}
where $\vec{\Sigma}_{\infty}$ is the frequency independent real part of the
self-energy~\cite{Schirmer1983}.  $\vec{\varepsilon}$ is a diagonal matrix with the
elements given by the zeroth-order state energies. By defining the matrix of spectral
functions in terms of the imaginary part of the single-particle Green function
($\vec{A}(\omega)=\frac1\pi|\mathrm{Im}\vec{G}(\omega)|$) and likewise for the spectral
function of the self-energy
($\vec{S}(\omega)=\frac1\pi|\mathrm{Im}\vec{\Sigma}(\omega)|$), and by the use of the 
superconvergence theorem~\cite{Altarelli1972} one can redefine the matrices in terms of
the frequency integrals:
\begin{eqnarray}
\vec{M}^{(n)}&=&\int_{-\infty}^{\infty}d\omega\,\omega^n \vec{A}(\omega),\quad
n=0\ldots2,\label{eq:Mn}\\
\vec{\Sigma}^{(0)}&=&\int_{-\infty}^{\infty}d\omega\,\vec{S}(\omega).\label{eq:S1}
\end{eqnarray}

\begin{figure}
\includegraphics[width=\columnwidth]{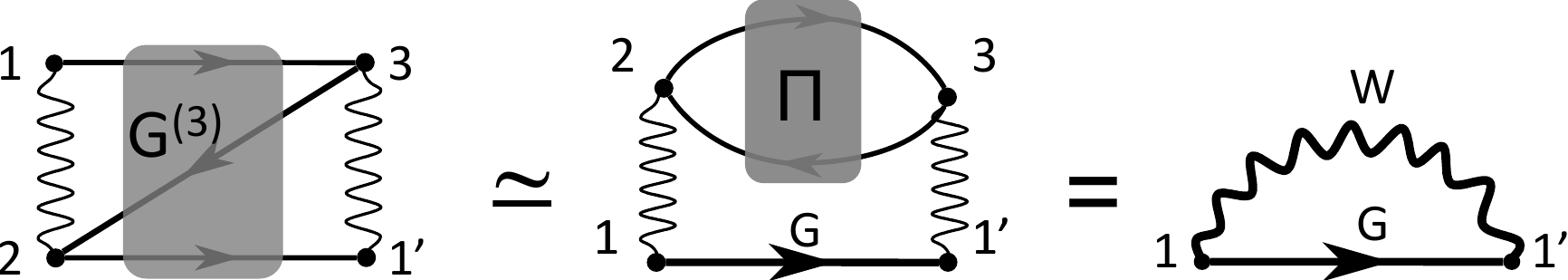}
\caption{Diagrams for the electron self-energy: The exact expression in terms of the $2p1h$
  response function ($G^{(3)}$) in which the entrance and the exit channels cannot be
  separated by cutting one fermion line can be  factorized approximately into the exact
  particle-hole ($p\mhyphen h$) propagator ($\Pi$) and the exact $1p$ Green's function ($G$)
  or, equivalently, into a product of the exact screened Coulomb interaction ($W$) and
 the  exact Green's function.
  \label{fig:diagram}}
\end{figure}

Equation~(\ref{eq:srule2}) expressed in the basis of Hartree-Fock states is central for
our discussion.  In order to be feasible it should be written in a finite basis and
coupled with an appropriate approximation for the self-energy operator. It is well known
that the self-energy can be recast in terms of the $2p1h$ ($2h1p$) six-point correlation
function~\cite{Winter1972,Schirmer1983}. Our approach corresponds to its factorization
into the product of $p\mhyphen h$ and $p$($h$) correlation functions
(Fig.~\ref{fig:diagram}) linking it to the established $GW$
approximation~\cite{Hedin,Onida2002}.  It is remarkable that after all complicated
Lehmann-type expressions for the correlation functions are substituted into the $\vec
\Sigma=i\vec G\vec W$ and the frequency integral in Eq.~(\ref{eq:S1}) is performed we obtain
an expression for the energy-uncertainty of a particularly simple form:
\begin{widetext}
\begin{eqnarray}
\vec{\sigma}^2_{\alpha\beta}=\vec{M}^{(2)}_{\alpha\beta}-[\vec{M}^{(1)}]_{\alpha\beta}^2
=\vec{\Sigma}^{(0)}_{\alpha\beta}&=&2\sum_i\sum_{j,k,k,k'}\langle\alpha
i|jk\rangle\langle\beta i|j'k'\rangle\left[
n_jn_{j'}\delta_{kj}\delta_{k'j'}+n_k(1-n_j)\delta_{kk'}\delta_{jj'}
-\rho_{k'j'}\rho_{kj}\right],
\label{eq:S0-GW}
\end{eqnarray}
\end{widetext}
where $\rho_{ij}=\langle 0|c_i^\dagger c_j|0\rangle$ is the single-particle density matrix
and the Coulomb matrix elements are defined as
$\langle\alpha\beta|\gamma\delta\rangle=\int d(\vec{r}_1\vec{r}_2)\phi_\alpha(\vec{r}_1)
\phi_\beta(\vec{r}_1)\phi_\gamma(\vec{r}_2)\phi_\delta(\vec{r}_2)/|\vec{r}_1-\vec{r}_2|$.
The prefactor 2 arises from the spin degeneracy. Apart from the factorization of the
six-point correlation function Eq.~(\ref{eq:S0-GW}) is exact. It is sufficient to
determine only the density matrix of the correlated ground state. For small molecules or
clusters this can be done at any desired level of accuracy by the configuration
interaction (CI) approach which, however, currently not feasible for
fullerenes. Therefore, we consider a further approximation widely known as $GW^0$
(Ref.~\onlinecite{SCGW5}) and treat the screened Coulomb interaction on the random phase
approximation level. This is equivalent to the configuration interaction singles (CIS)
treatment of the excited states. In view of the Brillouin's theorem~\cite{Brillouin1932}
the ground state density matrix is $\rho_{ij}=n_i\delta_{ij}$ leading to the
energy-uncertainty expressible in terms of the Coulomb matrix elements only:
\begin{equation}
\vec{\sigma}^2_{\alpha\beta}=2\sum_i\sum_{j,k}
\langle\alpha i|jk\rangle
\langle\beta i|jk\rangle n_k(1-n_j).
\label{eq:S0-GW0}
\end{equation}
The main computational burden is the transformation of the Coulomb matrix elements from
the atomic to molecular orbital basis which scales as $\mathcal{O}(N^5)$ in the brute
force implementation. To obtain convergent results we used very large basis sets (up to
6-311++G(3df,3pd)) resulting in $N=2340$ functions for C$_{60}$. Integral transformations
for this number of basis functions is not feasible with any standard quantum chemistry
package and required a parallelized implementation fully accounting for the symmetry of
the system.

Our calculations clearly indicate a peculiarity of SAMOs: a strong localization in the
energy domain (cf. $\vec \sigma_\mathrm{HOMO}=16.0$~eV and $\vec
\sigma_\mathrm{SAMO_d}=6.1$~eV) (Fig.~\ref{fig:c60}) or their extended life-time even in
comparison with the life-time of HOMO or LUMO states. This finding endorses the potential
of SAMOs as transport channels in molecular electronic devices, since the energy is hardly
dissipated during the short transport time.  Our results are also unexpected from the
Landau's theory of Fermi liquids~\cite{PinesNozieres} and illustrate that finite systems
possess electronic excitations that differ drastically from quasiparticles in extended
matter. Namely, a particle excitation may decay into two particles and one hole. This
process may recur for many generations or it may stop after a few.  For the former case a
large number of many-particle peaks form a Lorentzian envelope, the so-called
\emph{quasiparticle}. If the decay stops after a finite number of branchings, only Dirac
peaks appear in the single particle spectrum (denoting a localization in the Fock space).
A decisive discriminating factor was shown to be the particle's energy ($\epsilon$)
(Ref.~\onlinecite{Altshuler1997}). Our \emph{ab initio} approach suggests that it is
actually the kinetic energy $\epsilon_K$ that governs the decay. We find that even though
the energies of electrons with different radial character may be very similar their
kinetic energy is substantially different (as follows from the specific form of the
Kohn-Sham potential for this system, Ref.~\onlinecite{VKS}), which imposes strong
restrictions on the Coulomb matrix elements and leads to the localization of SAMOs in the
Fock space. This hints on the prolonged life-times and stipulates the observed high
stability of these states.

 We thank the DFG for financial support through SFB 762 and H. Petek for fruitful
discussions.  J.B. acknowledges financial support by Stanford Institute for Materials \&
Energy Science and Stanford Pulse Institute for ultrafast science where parts of this work
were completed. 

\end{document}